\documentclass[aps,showpacs,twocolumn]{revtex4-2}
\usepackage[utf8]{inputenc}
\usepackage{amsfonts}
\usepackage{amssymb}
\usepackage{amsmath}
\usepackage{graphicx}
\usepackage{epsfig}
\usepackage{subfigure}
\usepackage{appendix}
\usepackage{hyperref}
\usepackage{color}

\hypersetup{hypertex=true, colorlinks=true, linkcolor=blue, urlcolor=blue, citecolor=blue}

\begin{document}

\title{Dynamic destruction of magnetic order in a quantum Ising chain with
oscillating transverse field}
\author{E. S. Ma}
\author{Z. Song}
\email{songtc@nankai.edu.cn}
\affiliation{School of Physics, Nankai University, Tianjin 300071, China}
\begin{abstract}
We study the dynamic response of magnetic domain walls in low-lying excited
states of an Ising chain to an oscillating transverse field. Based on the
exact instantaneous eigenstates, we find that when the frequency of the
external field is in off-resonant regions, the domain wall exhibits Bloch
oscillation, maintaining the magnetic order. However, the magnetic order is
destroyed when the field is at resonant frequency. Numerical simulations of
the dynamics of magnetization and entanglement entropy for initial states
with single and double domain walls accord with the predictions. These
findings reveal the nontrivial effect of a monochromatic electromagnetic
field on quantum spin dynamics.
\end{abstract}

\maketitle

\section{Introduction}

\label{introduction}

In recent years, increasing attention has been drawn to periodically driven
systems~\cite{moessner2017equilibration, eckardt2017colloquium}, referred to
as Floquet systems, which can be described by a series of time-independent
Floquet states and energies that are analogous to the Brillouin-zone
artificial dimension~\cite{shirley1965solution}. Floquet systems could be a
promising platform to explore a variety of novel functionalities of a
quantum system, for example, time crystals~\cite{else2016floquet}, Floquet
maser~\cite{jiang2021floquet,liu2021masing}, Floquet Raman transition~\cite%
{shu2018observation}, prethermalization\cite{peng2021floquet}, Floquet
cavity electromagnonics \cite{xu2020floquet}, and Floquet polaritons~\cite%
{clark2019interacting}. A simplest Floquet system is a two-level atom driven
by a periodic oscillating field, which leads to phenomena like Rabi
oscillations. A natural question arises: what happens to an ensemble of such
two-level systems when they are coupled? Intuitively, the correlation
between atoms should affect the dynamics, which also depends on the initial
configuration. The simplest interaction between atoms is the
nearest-neighbor Ising-type coupling, which results in ferromagnetic domains
in low-lying and ground states. However, when a transverse field is applied,
the situation becomes complicated, even though the field is constant. Most
of studies focuses on the phase diagram of the ground state \cite%
{Sachdev1999quantum,li2009exact,zhang2015topological,zhang2021quantum}. From the perspective of low-lying excited states, it turns out that the magnetic domain wall
exhibits dynamic behavior characteristic of Bloch oscillations (BO).
Traditionally, BO describes the periodic motion of a wave packet subjected
to an external force in a lattice \cite{bloch1929quantenmechanik,
zener1934theory}. The related study mainly focus on the non-interacting
system. Therefore, as a non-equilibrium dynamic phenomenon in quantum
many-body systems, magnetic BOs in the quantum spin chains have attracted
much attention from researchers \cite{kyriakidis1998bloch, cai2011quantum,
shinkevich2012spectral, kosevich2013magnon, shinkevich2013numerical,
syljuaasen2015dynamical, hansen2022magnetic, zhang2024magnetic}. Notably,
inelastic neutron scattering experiments have provided evidence for the
existence of magnetic BOs in the magnetically identical material $\mathrm{%
CoCl_{2}\cdot 2D_{2}O}$ \cite{hansen2022magnetic}.

In this work, motivated by the question above, we study the dynamic response
of magnetic domain walls in the low-lying excited states of an Ising chain
to an oscillating transverse field. We show exactly that the instantaneous
spectrum for a single domain wall is equally spaced and time-independent. It
allows us to find that when the frequency of the external field is in
off-resonant regions, the domain wall exhibits Bloch oscillations,
maintaining the magnetic order. However, the magnetic order is destroyed
when the field is at resonant frequency. Numerical simulations of three
quantities, fidelity, magnetization, and entanglement entropy, in cases with
single and double domain walls accord with the predictions. This indicates
that a pre-engineered quantum system can be sensitive to the impact of a
driven field, resulting in the dynamic destruction of magnetic order within
a narrow frequency window. These findings reveal the nontrivial effect of a
monochromatic electromagnetic field on quantum spin dynamics and may benefit
the design of quantum devices.

This paper is organized as follows. In Sec. \ref{model}, we introduce the
Hamiltonian of the quantum Ising chain, which is driven by an oscillating
external field. The effective Hamiltonian for a single domain wall is
presented. In Sec. \ref{Dynamics of the domain wall}, we provide the exact
solution for the instantaneous eigenstates of the effective Hamiltonian. The
time evolution operator is obtained under the rotating wave approximation.
Sec. \ref{Dynamic demonstrations} is devoted to the numerical results of the
dynamics for the quantum spin chain. Finally, we conclude our findings in
Sec. \ref{summary}.

\begin{figure*}[tbh]
\centering
\includegraphics[width=0.8\textwidth]{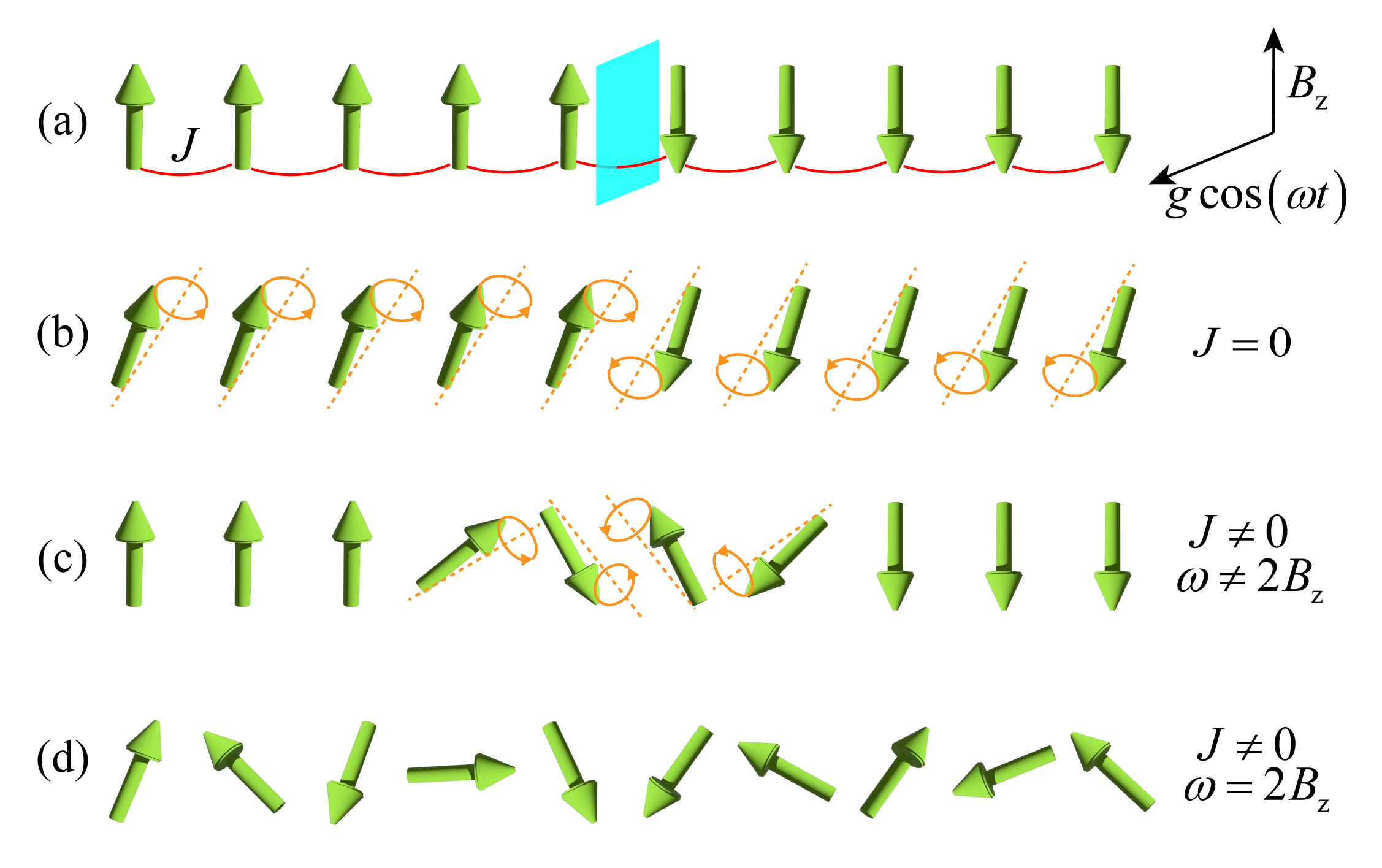}
\caption{Schematic illustrations of the system we studied and the main
results of this work. (a) A quantum chain with nearest-neighbor Ising-type
interaction of strength $J$. There are two external fields: one is a constant
field in the $z$-direction, while the other is a periodic field with
frequency $\protect\omega $ in the $x$-direction. The configuration of spins
represents an initial state for the following dynamic processes. This
initial state is a ferromagnetic state with a single domain wall at the
center. The main results of this paper are sketched in the following. (b) In
the case of decoupling, that is, when $J=0$, every spin in the same domain
evolves in the same phase, leaving the domain wall unchanged. (c) When the
driving field is off-resonance, the domain wall is slightly disturbed,
leaving it almost unchanged. (d) When the driving field is on-resonance, the
domain wall is destroyed, resulting in a disordered phase.}
\label{fig1}
\end{figure*}

\section{Model and effective Hamiltonian}

\label{model}

The model we consider is a quantum spin chain with a simple Ising
interaction, which is driven by an oscillating external field. The
Hamiltonian takes the form%
\begin{equation}
H=H_{0}+H_{\text{\textrm{D}}}(t),  \label{H_spin}
\end{equation}%
where 
\begin{equation}
H_{0}=-J\sum_{j=1}^{N-1}\sigma _{j}^{z}\sigma
_{j+1}^{z}-B_{z}\sum_{j=1}^{N}\sigma _{j}^{z}
\end{equation}%
represents a spin chain in longitudinal magnetic field $-B_{z}$ ($B_{z}>0$),
and with ferromagnetic Ising coupling $-J$. The transverse magnetic field
term 
\begin{equation}
H_{\text{\textrm{D}}}(t)=-g\cos (\omega t)\sum_{j=1}^{N}\sigma _{j}^{x}
\end{equation}%
is time dependent. Here $\sigma _{j}^{\alpha }$ ($\alpha =x,y,z$) are the
Pauli operators on site $j$. When $J=0$, the Hamiltonian $H$ describes a set
of decoupled spins. Then the physics of the whole system can be explained by
a single spin. Taking the rotating wave approximation (RWA) under the
condition $\omega +2B_{z}\gg \left\vert \omega -2B_{z}\right\vert $, any
given initial state of $j$-th spin $c_{j}^{+}\left( 0\right) \left\vert
\uparrow \right\rangle _{_{j}}+c_{j}^{-}\left( 0\right) \left\vert
\downarrow \right\rangle _{_{j}}$ evolves to a state $c_{j}^{+}\left( t\right)
\left\vert \uparrow \right\rangle _{_{j}}+c_{j}^{-}\left( t\right)
\left\vert \downarrow \right\rangle _{_{j}}$ at time $t$. It turns out that 
$c_{j}^{\pm }\left( t\right) $\ exhibits similar dynamical behavior when
driven by a time-independent Hamiltonian $-g\sigma^{x}/2$ at resonance,
where $\omega =2B_{z}$. A natural question arises: What happens to the
resonance in the case with nonzero $J$? On the other hand, when $B_{z}=0$,
the Hamiltonian $H$ reduces to a time-dependent transverse field Ising
chain. The instantaneous eigenstates can be obtained through the
Jordan-Wigner transformation \cite{pfeuty1970one} when a periodic boundary
condition is applied, and this serves as a unique paradigm for understanding
the quantum phase transition \cite{Sachdev1999quantum}. A nonzero longitudinal field term involves string
operators in the fermion representation, and thus the corresponding
Hamiltonian cannot be solved using traditional methods, such as the
Bogoliubov transformation. When $g=0$, the Hamiltonian $H$ can be exactly
solvable. This is crucial and serves as the starting point of the
investigation.

In this work, we focus on the dynamics within the low-energy subspace of the
model. Consequently, we opt for a perturbative approach to derive an
effective Hamiltonian that captures the low-energy dynamics. To achieve
this, we consider the weak field regime where $B_{z}$, $\left\vert
g\right\vert \ll J$, and we treat the transverse field term $H_{\text{%
\textrm{D}}}(t)$ as a perturbation in our subsequent analysis. It is worth
noting that the eigenstates of $H_{0}$ can all be expressed in a tensor
product form, with a set number of spins aligned either parallel or
antiparallel to the $z$ direction. The ground state of $H_{0}$ is given by $%
\left\vert \Uparrow \right\rangle =\prod_{j=1}^{N}\left\vert \uparrow
\right\rangle _{_{j}}$, with an energy $\mathcal{E}_{\mathrm{G}%
}=-N(J+B_{z})+J$. Our attention is directed towards the low-energy subspace $%
\{\left\vert \phi _{m}^{\pm }\right\rangle \}$, which is composed of states
featuring a single magnetic domain wall. In this context, $\left\vert \phi
_{m}^{\pm }\right\rangle $ denote two distinct types of domain-wall states 
\begin{equation}
\left\vert \phi _{m}^{+}\right\rangle =\prod_{l\leqslant m}\sigma
_{l}^{-}\left\vert \Uparrow \right\rangle ,\left\vert \phi
_{m}^{-}\right\rangle =\prod_{l>m}\sigma _{l}^{-}\left\vert \Uparrow
\right\rangle ,
\end{equation}%
where $\sigma _{j}^{-}=(\sigma _{j}^{x}-i\sigma _{j}^{y})/2$ is the lowering
operator, and $m=1,2,$ $...,$ $N-1$ indicates the spatial position of the
domain wall. We then have $H_{0}\left\vert \phi _{m}^{\pm }\right\rangle =%
\mathcal{E}_{m}^{\pm }\left\vert \phi _{m}^{\pm }\right\rangle $ with $%
\mathcal{E}_{m}^{\lambda }=$ $-N(J+\lambda B_{z})+3J+2\lambda mB_{z}$, with $%
\lambda =\pm $. The effective Hamiltonian can be derived by expressing the
Hamiltonian $H$ in the subspace spanned by the basis set $\left\{ \left\vert
\phi _{m}^{\pm }\right\rangle \right\} $ 
\begin{equation}
H_{\mathrm{eff}}=h_{+}+h_{-},  \label{H eff}
\end{equation}%
where%
\begin{eqnarray}
&&h_{\lambda }=-g\cos (\omega t)\sum_{m=1}^{N-2}\left( \left\vert \phi
_{m}^{\lambda }\right\rangle \left\langle \phi _{m+1}^{\lambda }\right\vert +%
\text{H.c.}\right)  \notag \\
&&+2\lambda B_{z}\sum_{m=1}^{N-1}m\left\vert \phi _{m}^{\lambda
}\right\rangle \left\langle \phi _{m}^{\lambda }\right\vert +\mathcal{E}%
_{0}^{\lambda }\left\vert \phi _{m}^{\lambda }\right\rangle \left\langle
\phi _{m}^{\lambda }\right\vert .
\end{eqnarray}%
We note that $H_{\mathrm{eff}}\ $consists of two independent parts, that is $%
\left[ h_{+},h_{-}\right] =0$. Each of them describes a single-particle
chain with a uniform hopping strength of $-g\cos (\omega t)$ and a uniformly
tilted potential with a slope $2B_{z}$. In the following, we only focus on
the dynamics in $\{\left\vert \phi _{m}^{+}\right\rangle \}$ subspace, and
denote $\left\vert m\right\rangle =\left\vert \phi _{m}^{+}\right\rangle $
for simplicity. The analysis is similar for that of $\{\left\vert \phi
_{m}^{-}\right\rangle \}$ subspace.

\section{Dynamics of the domain wall}

\label{Dynamics of the domain wall}

In this section, we will study the dynamics of the domain wall in the
framework of the effective Hamiltonian, given in Eq. (\ref{H eff}). In order
to obtain the analytical result, we study the system in the thermodynamic
limit, $N=\infty $. For simplicity, we neglect the constant term and reset
the origin of the site position. We rewrite one of two parts of $H_{\mathrm{%
eff}}$\ in the form

\begin{eqnarray}
h_{+} &=&-g\cos (\omega t)\sum\limits_{n=-\infty }^{+\infty }(|n\rangle
\langle n+1|+\text{\textrm{H.c}})  \notag \\
&&+\mathcal{\omega }_{0}\sum\limits_{n=-\infty }^{+\infty }n|n\rangle
\langle n|,  \label{h+}
\end{eqnarray}%
with $\mathcal{\omega }_{0}=2B_{z}$, which is the starting point of the
following discussion. First, we will present the exact solution for $h_{+}$
at a given time $t$. Second, based on the instantaneous eigenstates, we will
obtain the time evolution operator under the rotating wave approximation
(RWA). Third, we will analyze the results.

\subsection{Instantaneous Stark ladder}

\label{Instantaneous Stark ladder}

It is well known that the spectrum of $h_{+}$ is equally spaced when $g\cos
(\omega t)$\ is a real constant, supporting periodic dynamics with the
period $2\pi /\mathcal{\omega }_{0}$ \cite%
{hartmann2004dynamics,bloch1929quantenmechanik,zener1934theory}. Here we
briefly review the derivation of the solution, which are the instantaneous
eigenstates of the time-dependent of $h_{+}$. For given $t$, the eigenstate $%
{\left\vert \psi _{m}\right\rangle }$\ satisfying the Schr\"{o}dinger
equation 
\begin{equation}
{H\left\vert \psi _{m}\right\rangle =E_{m}\left\vert \psi _{m}\right\rangle ,%
}
\end{equation}%
of can always be written in the form 
\begin{equation}
{\left\vert \psi _{m}\right\rangle =\sum_{n}c_{nm}|n\rangle .}
\end{equation}%
The coefficient ${c_{nm}}$\ is determined by the equation 
\begin{equation}
{c_{n+1,m}+c_{n-1,m}=\frac{n\mathcal{\omega }_{0}-E_{m}}{g\cos (\omega t)}%
c_{nm},}
\end{equation}%
which accords with the recurrence relation of the Bessel function {$J_{n}$}
of argument $x$, i.e., 
\begin{equation}
{J_{n+1}(x)+J_{n-1}(x)=2\frac{n}{x}J_{n}(x).}
\end{equation}%
Then, we have 
\begin{equation}
{c_{nm}=J_{n-m}}\left[ {\frac{2g\cos (\omega t)}{\omega _{0}}}\right] {,}
\end{equation}%
and%
\begin{equation}
{E_{m}=m\omega _{0}.}
\end{equation}%
We can see that the eigenenergy ${m\mathcal{\omega }_{0}}$ is independent of
time $t$. Then the instantaneous eigenstate can be exactly expressed as ${%
\left\vert \psi _{m}\right\rangle =\sum_{n}J_{n-m}}\left[ {\frac{2g\cos
(\omega t)}{\omega _{0}}}\right] {|n\rangle }$. This is an important basis
for the following investigations.

\subsection{Floquet solution}

\label{Floquet solution}

The effective Hamiltonian $h_{+}$\ is a Floquet system.\ Now, we investigate
the dynamics of the system by the Floquet\ solution of the Schr\"{o}dinger
equation 
\begin{equation}
i\frac{\partial }{\partial t}|\Psi \left( t\right) \rangle =h_{+}(t)|\Psi
(t)\rangle ,
\end{equation}%
The solution of the above equation can always be written in the form 
\begin{equation}
|\Psi (t)\rangle =\sum_{m}a_{m}(t)e^{-im\mathcal{\omega }_{0}t}|\psi
_{m}\rangle ,
\end{equation}%
where the coefficient $a_{n}(t)$\ obeys the following equation: 
\begin{equation}
\frac{\partial a_{m}}{\partial t}=-\sum_{n}a_{n}e^{i\left( m-n\right) \omega
_{0}t}\left\langle \psi _{m}\right\vert \frac{\partial }{\partial t}|\psi
_{n}\rangle .
\end{equation}%
Submitting the expression of $|\psi _{n}(t)\rangle $, we have the equation
about the coefficient $a_{n}(t)$ 
\begin{equation}
\frac{\partial a_{n}}{\partial t}=\frac{g\omega \sin (\omega t)}{\omega _{0}}%
(a_{n-1}e^{i\mathcal{\omega }_{0}t}-a_{n+1}e^{-i\mathcal{\omega }_{0}t}).
\end{equation}%
Taking the rotating-wave approximation (RWA) under the condition $\left\vert 
\mathcal{\omega }-\mathcal{\omega }_{0}\right\vert \ll \mathcal{\omega }_{0}$%
, we have 
\begin{equation}
i\frac{\partial a_{n}}{\partial t}=-\frac{g\omega }{2\omega _{0}}%
[a_{n-1}e^{i(\mathcal{\omega }_{0}-\omega )t}+a_{n+1}e^{i(\omega -\mathcal{%
\omega }_{0})t}].
\end{equation}%
It is clear that such a system is equivalent to an infinite ring threaded
through a varying flux, which increases linearly with time $t$. It has been
shown that the dynamics of any local initial state are periodic with
frequency $\left\vert \omega -\mathcal{\omega }_{0}\right\vert $, obeying a
quantum version of Faraday's law, which states that a linearly varying flux
plays a similar role to a linear static field \cite{hu2013dynamics,
zhang2025bloch}.

\subsection{Critical dynamics}

\label{Critical dynamics}

The above analysis shows that a domain wall exhibits a periodic behavior for
finite values of $\omega -\mathcal{\omega }_{0}$.\ It indicates that the
domain wall cannot be destroyed by a periodic driven field, except the case
where $\omega $\ approaches $\mathcal{\omega }_{0}$. Now we focus on the
dynamics of the system at resonance $\mathcal{\omega }=\mathcal{\omega }_{0}$%
, in which the above equation becomes 
\begin{equation}
i\frac{\partial a_{n}}{\partial t}=-\frac{g}{2}(a_{n-1}+a_{n+1}).
\end{equation}%
The problem is reduced to that of a uniform tight-binding chain with the
Hamiltonian 
\begin{equation}
h_{\mathrm{chain}}=-\frac{g}{2}\sum\limits_{l=-\infty }^{+\infty }(|l\rangle
\langle l+1|+\text{\textrm{H.c}}).
\end{equation}%
It can be diagonalized in the form 
\begin{equation}
h_{\mathrm{chain}}=\sum\limits_{k\in (-\pi ,\pi )}\varepsilon _{k}|k\rangle
\langle k|,
\end{equation}%
with 
\begin{equation}
|k\rangle =\frac{1}{\sqrt{2\pi }}\sum\limits_{l}e^{ikl}|l\rangle ,
\end{equation}%
and the spectrum 
\begin{equation}
\varepsilon _{k}=-g\cos k.
\end{equation}%
The time evolution of any given $\left\{ a_{l}(0)\right\} $ is 
\begin{eqnarray}
\left\vert \Psi (t)\right\rangle &=&\sum\limits_{l}a_{l}(t)|l\rangle  \notag
\\
&=&\sum\limits_{l}a_{l}(0)\sum_{n}i^{n-l}J_{n-l}(gt)|n\rangle .
\end{eqnarray}%
Here we consider a simple case where $g=1$ and $\left\vert \Psi
(0)\right\rangle =|0\rangle $. In this case, the straightforward derivation
yields 
\begin{equation}
a_{n}(t)=i^{n}J_{n}(t),
\end{equation}%
which ensures the probability distribution of the energy levels 
\begin{equation}
p_{n}(t)=\left\vert a_{n}(t)\right\vert ^{2}=\left\vert J_{n}(t)\right\vert
^{2}.
\end{equation}%
The estimation of the profile of $p_{n}(t)$ can be approached as follows.
The characteristic of the Bessel function reveals that a maximum occurs at
the boundary of $p_{n}(t)$, which can be interpreted as the wave front of
the expanding occupied energy levels. The position of this wave front,
denoted by $n_{c}$, can subsequently be identified using the equation 
\begin{equation}
{\frac{\partial p_{n_{c}}(t)}{\partial t}=0.}
\end{equation}%
Given the relationships of Bessel functions 
\begin{eqnarray}
J_{n_{c}-1}(t)-J_{n_{c}+1}(t) &=&2J_{n_{c}}^{\prime }(t),  \notag \\
J_{n_{c}+1}\left( t\right) +J_{n_{c}-1}\left( t\right) &=&2\left\vert
n_{c}/t\right\vert J_{n_{c}}\left( t\right) ,
\end{eqnarray}%
it follows that%
\begin{equation}
{2J_{n_{c}+1}(t)=2|n_{c}/t|J_{n_{c}}(t).}
\end{equation}%
When $n$ and $t$ are large, we find that ${J_{n_{c}+1}(t)\approx J_{n_{c}}(t)%
}$, leading to the approximation ${|n_{c}|\approx t}$. From this, we deduce
that the speed of the spreading of occupied energy levels is uniform.

\begin{figure*}[tbh]
	\centering
	\includegraphics[width=1.0\textwidth]{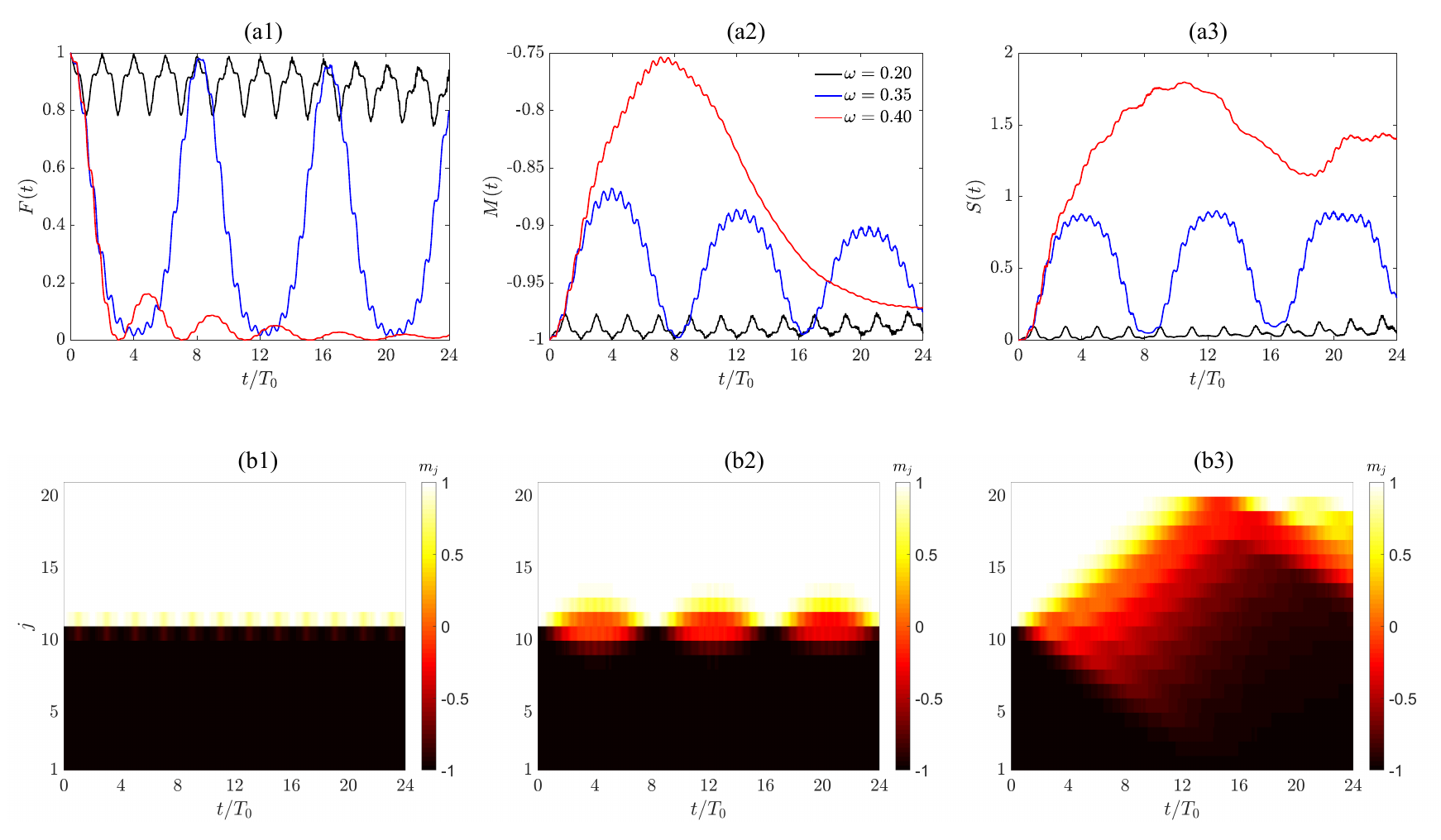}
	\caption{Plots of $F(t)$, $M(t)$, and $S(t)$,
		defined in Eqs. (\protect\ref{F(t)}), (\protect\ref{M(t)}) and (\protect\ref%
		{S(t)}), respectively, for the evolved state in Eq. (\protect\ref{onewall}),
		obtained by numerical solution of the Schr\"{o}dinger equation via the fourth-order
		Runge-Kutta method for a finite-size chain with several typical values of
		frequency $\mathcal{\protect\omega}$. Here, $%
		T_{0}=\protect\pi/B_{z}$ is the resonance period. (b1), (b2), and (b3) are
		the corresponding plots of {$m_{j}(t)$}, given in Eqs. (\protect\ref{m_j(t)}%
		), corresponding to three typical values of frequency $\omega=0.2$, $\omega=0.35$ and $\omega=0.4$, respectively. The other parameters are $J=1, g=0.05, B_{z}=0.2$ and $%
		N=20$. We can see that the patterns in (b1), (b2), and (b3) clearly reflect
		the impact of the oscillating field and provide an intuitive picture for
		understanding the curves in (a1), (a2), and (a3). The results are in
		accordance with our predictions.}
	\label{fig2}
\end{figure*}

\begin{figure*}[tbh]
	\centering
	\includegraphics[width=1.0\textwidth]{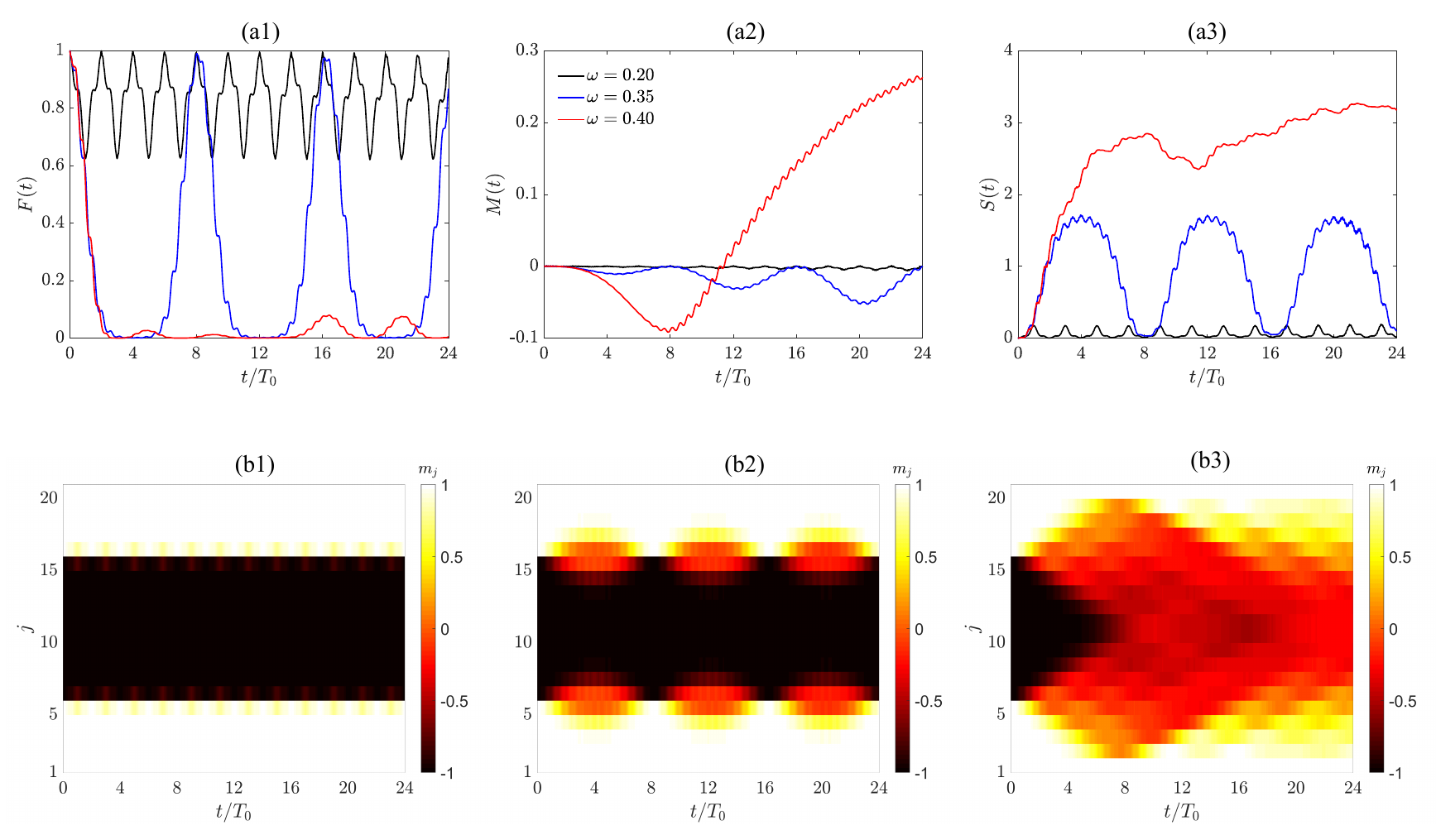}
	\caption{The same plots as those in Fig. \protect\ref{fig2}, obtained with the same parameters but for the initial state with double domain walls given in Eq. (\ref{twowalls}), are consistent with our predictions.}
	\label{fig3}
\end{figure*}

\section{Dynamic demonstrations}

\label{Dynamic demonstrations}

In this section, we investigate the dynamics in the quantum spin chain
through an analytical analysis of the Floquet effective Hamiltonian $h_{+}$,
given in Eq. (\ref{h+}). We will focus on the dynamic response of 
domain wall states to the oscillating transverse field. From the
analysis in the previous section, we note that the instantaneous eigenstates
are locally distributed and the instantaneous spectrum remains unchanged,
with equal spacing between energy levels, in the single domain wall subspace.

We consider a simple initial state 
\begin{equation}
\left\vert \Psi (0)\right\rangle =\prod_{l\leqslant N/2}\sigma
_{l}^{-}\left\vert \Uparrow \right\rangle
,
\end{equation}%
which is a single domain wall state. The initial state can also be expressed
as 
\begin{equation}
\left\vert \Psi (0)\right\rangle =\sum_{m}f_{m}\left\vert \psi
_{m}\right\rangle ,
\end{equation}%
where $f_{m}$\ is nonzero when $m$ is around zero under the condition of
finite $\left\vert \omega -\mathcal{\omega }_{0}\right\vert $. This ensures
the relation%
\begin{equation}
\left\vert \Psi (t+T)\right\rangle =\left\vert \Psi (t)\right\rangle ,
\end{equation}%
which indicates that the domain wall oscillates around its initial position,
rather than be destroyed. Here, the period $T=2\pi /\left\vert \omega -%
\mathcal{\omega }_{0}\right\vert $ tends to infinity at the resonance.
However, the wave front of the domain wall should be reflected at the
boundary for finite $N$. Then the domain wall becomes nonlocal after a long
time when $\omega \approx \mathcal{\omega }_{0}$.

We employ three quantities to demonstrate the results, which can also be the
observables in the experiment. Some predictions are given below based on the
effective Hamiltonian. (i) The fidelity is defined as%
\begin{equation}
F(t)=\left\vert \langle \Psi (0)\left\vert \Psi (t)\right\rangle \right\vert
^{2},  \label{F(t)}
\end{equation}%
which is periodic, that is, $F(t)=F(t+T)$ in the off-resonance regime,
according to the effective Hamiltonian. The fidelity should decay at
resonance.\ (ii) The magnetization is defined by%
\begin{equation}
m_{j}(t)=\left\langle \Psi (t)\right\vert \sigma _{j}^{z}\left\vert \Psi
(t)\right\rangle ,  \label{m_j(t)}
\end{equation}%
which remains almost unchanged in the off-resonance regime, according to the
effective Hamiltonian. We should have $m_{j}(t)\approx 0$\ after a
sufficient long time at resonance conditions. On the other hand, we also
introduce the average magnetization on the half chain, which is defined by%
\begin{equation}
M(t)=\frac{2}{N}\sum\limits_{j=1}^{N/2}m_{j}(t).  \label{M(t)}
\end{equation}%
It remains around $-1.0$ in the off-resonance regime but changes rapidly at
resonance, according to the effective Hamiltonian. (iii) The bipartite Von
Neumann entropy is defined by

\begin{equation}
S(t)=-\mathrm{Tr}\left( \rho _{\mathrm{A}}\ln \rho _{\mathrm{A}}\right) ,
\label{S(t)}
\end{equation}%
where%
\begin{equation}
\rho _{\mathrm{A}}=\text{\textrm{Tr}}_{\text{\textrm{B}}}\left( \left\vert
\Psi \left( t\right) \right\rangle \left\langle \Psi \left( t\right)
\right\vert \right) ,
\end{equation}%
\ is reduced density matrix for sublattice A. In this work, A and B denote
odd- and even-site sublattices, respectively. According to the effective
Hamiltonian, it remains almost unchanged in the off-resonance regime. In
contrast, $S(t)$ should increase over time at resonance.

To demonstrate and verify our predictions, we perform numerical simulations
for the three quantities. The time evolution 
\begin{equation}
\left\vert \Psi (t)\right\rangle =\exp (-iHt)\prod_{l\leqslant N/2}\sigma
_{l}^{-}\left\vert \Uparrow \right\rangle  \label{onewall}
\end{equation}%
is computed with the fourth-order
Runge-Kutta method for a finite spin chain \cite{ZINGG1999227}. The results are
presented in Fig. \ref{fig2}, and other parameters of the system are
presented in the caption. The results in Fig. \ref{fig2} indicate the
following features. (i) The fidelity $F(t)$ is periodic in the
off-resonance cases, and the periods increase as $\omega $ closes to $%
\mathcal{\omega }_{0}$. The fidelity exhibits oscillatory decay at the
resonance $\omega =\mathcal{\omega }_{0}$. (ii) The average magnetization
on the half-chain $M(t)$\ oscillates around $-1.0$ with a small amplitude when 
$\omega $ is far from $\mathcal{\omega }_{0}$. The amplitude increases as $%
\omega $ approaches $\mathcal{\omega }_{0}$. (iii) The bipartite Von
Neumann entropy $S(t)$ oscillates around a small positive number with a small
amplitude when $\omega $ is far from $\mathcal{\omega }_{0}$. The
amplitude increases as $\omega $\ approaches $\mathcal{\omega }_{0}$. (iv)
The distribution of magnetization $m_{j}(t)$ provides a clear picture for
the dynamics which illustrates the behaviors of $F(t)$, $M(t)$ and $S(t)$,
respectively. We note that this simulation is about finite-size chain. It is
presumably the case that the three quantities, $F(t)$, $M(t)$ and $S(t)$
are not periodic for the system at resonance in the large $N$ limit.

In addition, we also investigate the case for the initial state with two
domain walls, given by%
\begin{equation}
\left\vert \Psi (t)\right\rangle =\exp (-iHt)\prod_{N/4\leqslant l\leqslant
3N/4}\sigma _{l}^{-}\left\vert \Uparrow \right\rangle .  \label{twowalls}
\end{equation}%
In the framework of the effective Hamiltonian, we can predict that three
quantities, $F(t)$,\ $M(t)$\ and $S(t)$ are periodic when the distance
between the domain walls is sufficiently large and $\omega $\ is far from $%
\mathcal{\omega }_{0}$.\ The results are presented in Fig. \ref{fig3}, and
show that three quantities, $F(t)$,\ $M(t)$\ and $S(t)$ exhibit the similar
behaviors with that in Fig. \ref{fig2}.\ These numerical results are in
accordance with the analyses.

\section{Summary}

\label{summary}

In summary, we demonstrate the existence of a resonant window in a quantum
Ising chain for the frequency of the oscillating transverse field. The
underlying mechanism is based on two facts: (i) There are magnetic BOs in a
time-dependent transverse field; (ii) The Stark ladder is independent of the
perturbation on the transverse field. It allows us to describe the dynamics
of a single domain wall analytically in the effective Hamiltonian. We have
shown that when the frequency of the external field is in off-resonant
regions, the domain wall exhibits periodic fluctuations in a local region,
maintaining the magnetic order. Our main result is that the oscillation period approaches infinity at resonance, thereby destroying the
magnetic order. Our results, on the one hand, reveal the effect of the Ising
interaction on the Rabi oscillations. On the other hand, they may offer a
route to applications in quantum device engineering.

\acknowledgments This work was supported by the National Natural Science
Foundation of China (under Grant No. 12374461).

\section*{Data availability}

The data that support the findings of this article are openly
available \cite{ma_2025_17309070}.

%\bibliography{Floquet_Isingreference_1}

\end{document}